\documentclass[slac_one]{revtex4}
\usepackage{graphicx}
\usepackage{fancyhdr}
\def\Bd{\begin{math}B^0\end{math}}

\def\Bs{\begin{math}B^0_\mathrm{s}\end{math}}

\def\BsJpsiphi{\begin{math}B^0_\mathrm{s} \rightarrow J/\psi \phi \end{math}}
\def\Bsmumu{\begin{math}B^0_\mathrm{s} \rightarrow \mu \mu \end{math}}
\def\Bsphigamma{\begin{math}B^0_\mathrm{s} \rightarrow \phi \gamma \end{math}}

\def\Afb{\begin{math}A_\mathrm{FB} ( B^0 \rightarrow K^\mathrm{*} \mu \mu )\end{math}}
\def\CP{\begin{math}CP\end{math}}
\def\cms{\begin{math}\mathrm{cm}^{-2}\mathrm{s}^{-1}\end{math}}
\def\fb{fb\begin{math}^{-1}\end{math}}
\def\g{\begin{math}\gamma\end{math}}
\def\L0{Level-0}
\def\phis{\begin{math}\phi_\mathrm{s}\end{math}}

\def\SM{Standard Model}

\begin{document}

\title{LHCb prospects for full energy and beyond (incl. upgrades)} 

\author{R. Le Gac on behalf of the LHCb Collaboration}
\affiliation{CPPM, Aix-Marseille Universit\'e, CNRS/IN2P3, Marseille, France}

\begin{abstract}
The LHCb experiment is running at the Large Hadron Collider
to study \CP\ violation and rare decays in the beauty and charm sectors.
The physics potential is given for five key observables sensitive to new physics
in nominal conditions when $\sqrt s=14 ~\mathrm{TeV}$ and $\int \mathcal{L} = 2~\mathrm{fb}^{-1}$. 
The motivation and the strategy of the upgrade
envisaged for 2016 is presented with the expected performance for an integrated
luminosity of 50~\fb .
\end{abstract}

\maketitle

\thispagestyle{fancy}

\section{INTRODUCTION}
Studies of \CP\ violation and more generally flavored changing neutral
currents were essential in establishing the \SM , as the theory
describing the \CP\ violation observed in the laboratory. 
Today, they appear as a powerful tool to reveal
processes beyond the \SM\ and to understand their nature. In this
context, LHCb will play a key role.

\section{THE LHC$\mathrm{b}$ DETECTOR}
\begin{figure}[htb]
\centering
\includegraphics[height=10cm]{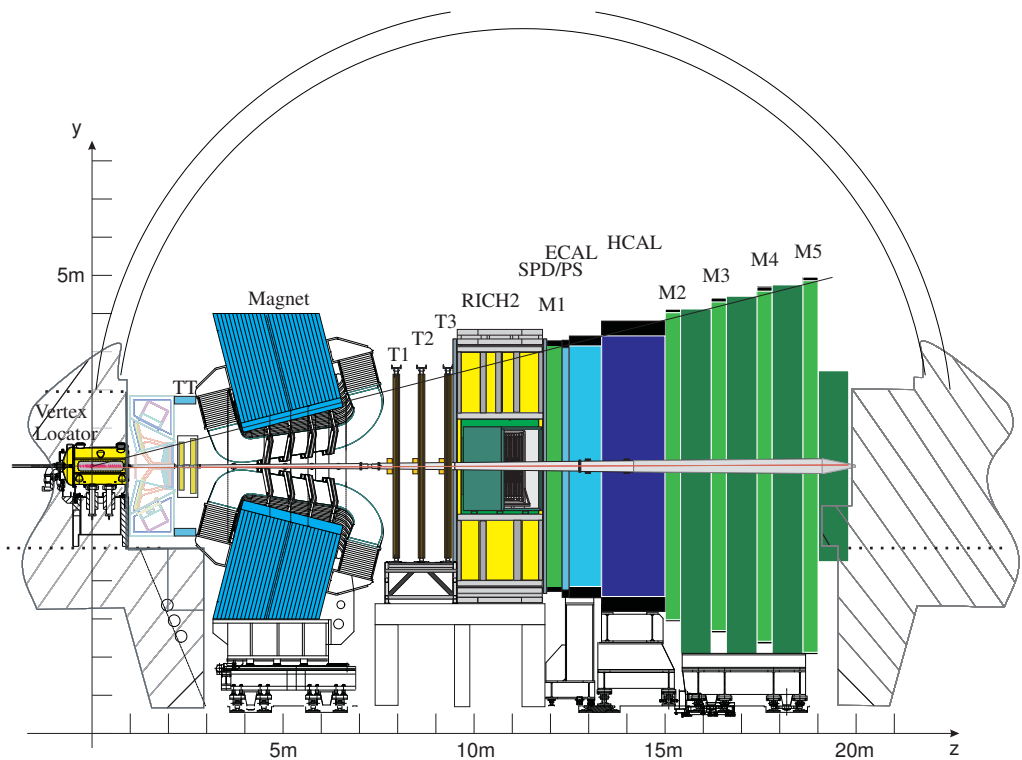}
\caption{Vertical view of the LHCb detector.}
\label{fig:detector}
\end{figure}

The LHCb detector~\cite{Alves:2008zz}, shown in Figure~\ref{fig:detector}, is a single-arm forward
spectrometer covering the forward region of the proton-proton interaction.
The detector geometry is driven by the kinematics of the $b \overline{b}$ pair production at the
LHC energy where both $b$ and $\overline{b}$ quarks mainly fly in the forward or
backward direction. The $b \overline{b}$ and the inelastic cross-sections are
estimated to 500 $\mu$b and 80 mb
respectively for proton-proton collisions at $\sqrt s=14 ~\mathrm{TeV}$.

The interaction point is surrounded by a silicon vertex detector. It measures
precisely the position of primary and secondary vertices as well as impact
parameters.
The tracking system is  composed of a magnet, deflecting particles in the
horizontal plane and two groups of tracking stations: the TT stations before 
the magnet and the T stations after.
On both sides of the tracking system, Ring Imaging Cerenkov counters (RICH)
are used to identify charged particles.
Further downstream, an electromagnetic calorimeter (ECAL) is used for 
photon detection and electron identification, followed by a hadronic calorimeter
and a muon detector. The latter is composed of five stations interleaved with iron shield.

The detector is designed to run at an instantaneous luminosity of 
$2 \times 10^{32}$~\cms\ and to accumulate an integrated luminosity of 2~\fb\ per year.

LHCb will collect from 5 to 6~\fb\ by end of 2015. 
The data will be recorded at different centre of mass energy: $\sqrt s=7 ~\mathrm{TeV}$
in 2010 -- 11 and at $\sqrt s \sim 14 ~\mathrm{TeV}$ in 2013 -- 15.
Nominal running conditions will be reached beginning 2014 with 
$\sqrt s=14 ~\mathrm{TeV}$ and an integrated luminosity of 2~\fb\ per year.

\section{PERFORMANCE IN NOMINAL CONDITIONS}

Among many observables accessible at LHCb, five key measurements have been
selected for their potential to reveal physics beyond the \SM ~\cite{Adeva:2009}:

\begin{itemize}
 \item the branching ratio of the \Bsmumu\ probing new mass and new coupling, 
 \item the weak phase $\phi_\mathrm{s} \sim 2 \times  \arg  V_\mathrm{ts}$ 
  in the \BsJpsiphi\ decay which probes new phases,
 \item the forward backward asymmetry, \Afb , and the photon polarization, $\mathrm{sin} 2 \psi$
 in \Bsphigamma\ which probe the Lorentz structure of the new coupling.
 The photon polarization is defined as
      $ \tan \psi \equiv 
	\left | 
	A_R \left ( B^0_\mathrm{s} \rightarrow \phi \gamma_R \right ) /
	A_L \left ( B^0_\mathrm{s} \rightarrow \phi \gamma_L \right )
	\right |$.
	In the \SM\ the b-meson decays  predominantly into a left handed photon
	due to the $V-A$ coupling of the $W$ boson.

 \item the weak phase $\gamma \sim \arg V_\mathrm{ub}$ in \Bd\ and \Bs\ decays
 mediated by two tree amplitudes since it is a reference measurement, not affected by new physics.
\end{itemize}

\begin{table}[htb]
\begin{center}
\caption{LHCb Sensitivity for $\sqrt s=14 ~\mathrm{TeV}$ and
$\int \mathcal{L} = 2~\mathrm{fb}^{-1}$ \cite{Adeva:2009}.}
\begin{tabular}{|l|c|c|} \hline  
\textbf{Observable}  & \textbf{Sensitivity}     & \textbf{SM prediction} \\ \hline
Br(\Bsmumu )         &  $1 \times 10^{-9}$ & $\left ( 3.35 \pm 0.32 \right ) \times 10^{-9}$  \cite{Buras:2006}\\ \hline
\phis\               &  0.03 rad                & $\left (0.0363\pm 0.0017\right)$ rad~\cite{CKM} \\ \hline
zero of \Afb\        & 0.5 GeV$^2$              & 4.36 GeV$^2$~\cite{Beneje:2005} \\ \hline
$\mathrm{sin} 2 \psi$ in \Bsphigamma\ &  0.22        & 0.10~\cite{Grinstein}   \\ \hline
\g\ (tree amplitude) & $\sim 5$ deg             & $\left(67.2 \pm 3.9 \right)$ deg~\cite{CKM}      \\ \hline
\end{tabular}
\label{table1}
\end{center}
\end{table}

The summary of the expected sensitivity for one year of data taking under nominal conditions is
given in Table~\ref{table1}.
The branching ratio \Bsmumu\ can be observed at 3 standard deviations
with respect to the \SM\ prediction.
The sensitivity to the weak phase \phis\ is comparable to the \SM\ expectation.
The sensitivity to the week phase \g\ is ten time better than the expected value,
improving the accuracy of the current measurement by a factor $\sim 4$.
The zero crossing point of \Afb\ can be determined with an accuracy 
ten time better than the \SM\ expectation, giving a good sensitivity to discriminate
between different models of new physics~\cite{Ali:1999mm, Altmannshofer:2008dz}.
Finally the photon polarization in \Bsphigamma\ 
can be determined with an accuracy close to the \SM\ value.
Improving the performance far beyond that point requires an upgrade of the LHCb detector.

\section{The LHC$\mathrm{b}$ DETECTOR UPGRADE}
New physics will be hopefully discovered by Atlas/CMS and LHCb in the coming years.
Better accuracies as well as more observables will be required by the end 2015
to understand its nature and to elucidate its flavor structure.
On the contrary, if no new physics hints are discovered, more statistics will be required, to probe
higher mass scales.

In this context, the aim of the LHCb upgrade is to gain a factor 10 in 
the instantaneous luminosity, to improve the trigger efficiency for hadronic channels
and to collect at least 50~\fb .

The challenge of this update is related to the number of 
proton-proton interactions per beam beam crossing.
The average number of interactions per crossing is equal to 0.4 for the nominal conditions
at $2\times10^{32}$~\cms .
It increases to 2.3 at $1\times10^{33}$~\cms and reaches 4 interactions per crossing at $2\times10^{33}$~\cms . 
In the latter case, all crossings have at least one interaction.
These extreme conditions put heavy requirements on the tracking as well as on the trigger algorithms 
since the number of tracks is roughly multiplied by a factor 10 and the number of vertices by a factor 4 
with respect to those expected at nominal conditions.

\subsection{New trigger strategy}

The current LHCb trigger contains two stages,
the \L0\ and the High Level Trigger (HLT).
The \L0\ reduces the rate from 40~MHz down to 1~MHz. 
It is based on custom electronics receiving dedicated information 
from the calorimeters and from the muon detector. 
It looks for lepton and hadron candidates with a high transverse
momentum.

The detector is read out at 1~MHz.
The HLT trigger reduces the rate down to 2~kHz.
The high level trigger is a software trigger running on a dedicated CPU farm and
receiving the full detector information. By running tracking and vertexing
algorithms it confirms the \L0\ candidates and select leptons and hadron with a
high transverse momentum as well as a high impact parameters. 
More elaborate algorithms, close to the off-line selections are then applied to select
inclusive or exclusive b-meson decays.
   
\begin{figure}[htb]
\centering
\includegraphics[height=7.0cm]{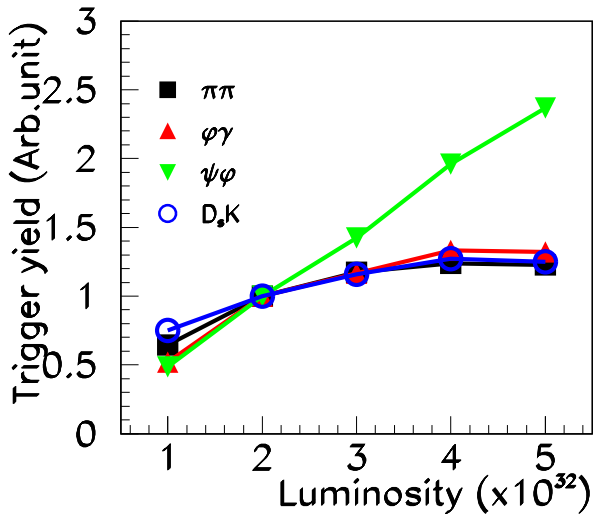}
\caption{The trigger yield as a function of the luminosity for different decays of b-mesons.
Each point is normalized to the trigger yield expected in nominal conditions 
for an integrated luminosity of 2~fb$^{-1}$. 
The different curves show the \L0\ saturation when increasing the luminosity
for final state with hadrons.}
\label{fig:lumi}
\end{figure}

As shown in Figure~\ref{fig:lumi}, the \L0\ saturates for hadronic channels when the luminosity
increases.
At high luminosity we cannot rely only on the transverse momentum cut for efficient triggering. 
A software based trigger, however, allows use of many discriminants including track impact parameters and combinations of different criteria.

The easiest way to implement this new scheme
is to readout the  whole detector at 40~MHz and to trigger on
the interesting events in the HLT. The HLT output rate will also be
increased by a factor 10, from 2 to 20~kHz.

This software trigger is very flexible and can be adapted to any kind of
scenarii. It will be highly selective on b-meson decays providing a reduction
factor on minimum bias events at the level of $10^5$.

As a consequence, the front-end electronics have to be rebuild in order to cope
with the 40~MHz readout.
This is also the case of all very front-end ASICs which are limited to a 
readout frequency of 1~MHz.
In addition, increasing the instantaneous luminosity by a factor 10 means higher
occupancy and higher radiation dose for the detector around the beam pipe.
Therefore, the inner parts of the tracking system have to be redesigned, namely
the vertex detector, the trigger tracker TT and the inner tracker IT.

\subsection{Upgrading the vertex detector}
The current vertex detector is a silicon strip detector with $r$ and $\phi$
geometry. The strip pitch varies between 35 and 100 $\mu$m.
It will be replaced by a pixel detector providing a very low occupancy for each
channel, reducing the combinatorial for the tracking algorithm. 
The base line is a modified version of the TimePix readout chip developed by the Medipix
Collaboration at CERN~\cite{Medipix}. The chip is a matrix of $256 \times 256$ pixels. 
Each pixel has a square shape with a size of $55 \times 55 ~\mu\mathrm{m}^2$.

\subsection{Upgrading the tracking system}
The trigger tracker and the inner tracker are silicon micro-strip detectors.
The strip pitch is 200 micron and the strip length varies between 11 and 33 cm.
Two options are envisaged to replace them.
The first one is a silicon option where new modules are built identical to
the existing ones but with a new rad-hard front chip which can be read at 40~MHz.
The second option is a fiber tracker based on five layers of scintillating
fibers with a fiber diameter of 250~$\mu$m. In such system the light collection
is performed via silicon photomultiplier and all the front-end electronics is
pushed outside the detector acceptance, to avoid the use of rad-hard
electronics.

The outer tracker (OT) is a gaseous detector based on 5~mm diameter
and very long straw tubes. Their length
varies between 2.4 and 5~m.
No replacement is foreseen in the OT detector in the 2016 upgrade.
Simulation shows that the occupancy is too high when running at an instantaneous
luminosity of $2 \times 10^{33}$~\cms . 
The instantaneous luminosity will be therefore limited to $1 \times 10^{33}$~\cms .

\subsection{Upgrading the RICH photon detectors}
The RICH photon detector is an HPD.
The anode is a matrix of $32 \times 32$ pixels.
The size of the pixels at the entrance window is a square of 
$2.5 \times 2.5~\mathrm{mm}^2$ and the readout frequency is limited to 1~MHz.
The replacement candidate is a multi-anode photo multiplier (MaPM) from Hamamatsu
which can be readout at 40~MHz.
The anode is a matrix of $8 \times 8$ pixels with a pixel size of $2 \times 2~\mathrm{mm}^2$.
In addition, the rectangular shape of the MaPM allows to reduce the dead
zone between photon detectors.

\section{Expected performance}
The estimated sensitivities for the 5 key observables are given in
Table~\ref{table2} for an integrated luminosity of 50~\fb , taking into
account all necessary trigger improvements in order to increase the hadronic channels yield.

\begin{table}[htb]
\begin{center}
\caption{
  LHCb Sensitivity with an upgraded detector
  for $\sqrt s=14 ~\mathrm{TeV}$ and
  $\int \mathcal{L} = 50~\mathrm{fb}^{-1}$.}
\begin{tabular}{|l|c|c|} \hline  
\textbf{Observable}  & \textbf{Sensitivity}     & \textbf{SM prediction} \\ \hline
Br(\Bsmumu )         &  $0.5\times10^{-9}$ & $\left ( 3.35 \pm 0.32 \right ) \times 10^{-9}$  \cite{Buras:2006}\\ \hline
\phis\               &  0.004 rad                & $\left(0.0363 \pm 0.0017\right)$ rad~\cite{CKM}   \\ \hline
zero of \Afb\        & 0.1 GeV$^2$              & 4.36 GeV$^2$~\cite{Beneje:2005} \\ \hline
$\mathrm{sin} 2 \psi$ in \Bsphigamma\ &  0.03        & 0.10                   \\ \hline
\g\ ($B \rightarrow DK$) & $< 1.4$ deg             & $\left(67.2\pm 3.9\right) $ deg~\cite{CKM}      \\ \hline
\end{tabular}
\label{table2}
\end{center}
\end{table}

The branching ratio \Bsmumu\ can be observed with a sensitivity corresponding 
to 15\% of the \SM\ predictions.
The sensitivity to the weak phase \phis\ is ten time better than the \SM\ expectation.
The sensitivity to the weak phase \g\ will be close to 1 degree.
The weak phase $\beta \sim \arg V_{tb}$ determines in 
$B^0 \rightarrow J/\psi K^0_\mathrm{S}$ and the weak pahse \g\ will therefore 
be measured with comparable accuracies, providing additional lever arm to constraint the unitarity triangle.
The sensitivity on the zero crossing point of \Afb\ corresponds to 2\% of the \SM\ expectation, and
provides an unprecedented power to discriminate between different models of 
new physics~\cite{Ali:1999mm, Altmannshofer:2008dz}.
Finally the photon polarization in \Bsphigamma\ can be measured at 3 
standard deviations at the \SM\ prediction.

Many more observables will be accessible with a hight
statistics, like the measurement of \phis\ in the \Bs\ decay to $\phi \phi$, the
measurement of \g\ in decays mediated by loop amplitudes or 
the fraction of longitudinal $K^\mathrm{*}$ polarization, $F_L$ in
$B^0 \rightarrow K^\mathrm{*} \mu \mu$.

A longer term upgrade is under studiy to run at $2 \times 10^{33}$, requiring 
a substantial R\&D effort. 
The outer tracker might be replaced by a thick fiber tracker.
The diameter of the fiber is 1~mm, they are group four by four and connected to position sensitive sensor like a PMT or silicon photomultiplier.
A new detector name TORCH~\cite{Charles:2010at} 
might be added between the RICH2 and the electromagnetic calorimeter. 
It could measure the time-of-flight of Cerenkov photons generated in a thin quartz plate. 
This novel detector could be used to identify hadrons in the low momentum range between 1 and 10~GeV.
Finally, we may improve the granularity of the inner part of the electromagnetic calorimeter.

\section{CONCLUSIONS}
First data collected by LHCb are very promising and confirm the expected detector performance.
LHCb can measure many observables probing physics beyond the \SM\ in
several different ways.
Many significant measurements will be achieved in the $b$ and also in the $c$ quark sector.

A challenging upgrade is under preparation, aiming at an integrated luminosity
of 100~fb$^{-1}$, improving efficiencies particularely on the hadronic channels. 
A letter of intent will be published end 2010.


\end{document}